\documentclass[12pt]{article}
\pdfoutput=1
\usepackage{graphicx} 
\usepackage{epsfig}
\usepackage{amsfonts}
\usepackage{amssymb}
\usepackage{xcolor}
\usepackage{mathrsfs}
\usepackage{amsmath}
\usepackage{upgreek}
\usepackage{graphicx}
\usepackage{cite}
\usepackage{calligra}
\usepackage{caption}
\usepackage{subcaption}
\usepackage{cite}
\usepackage{calligra}
\usepackage[english]{babel} 
\usepackage{hyperref}
\usepackage{stackengine}
\allowdisplaybreaks
\newcommand{\nc}{\newcommand}
\nc{\beq}{\begin{equation}} \nc{\eeq}{\end{equation}}
\nc{\beqa}{\begin{eqnarray}} \nc{\eeqa}{\end{eqnarray}}
\nc{\ba}{\begin{array}} \nc{\ea}{\end{array}}

\title{subleding_effecive_potential}

\begin{document}

\begin{center}

{\bf \Large\ Effective Potential in Subleading Logarithmic\\[0.1cm] Approximation in Arbitrary Non-renormalizable\\[0.2cm] Scalar Field Theory} \vspace{1.0cm}

{\bf \large R.M. Iakhibbaev$^{1}$, D.I. Kazakov$^{1}$,  A.I. Mukhaeva$^{1}$ \\[0.3cm] and D. M. Tolkachev$^{1,3}$}

\vspace{0.5cm}
{\it $^1$Bogoliubov Laboratory of Theoretical Physics, Joint Institute for Nuclear Research, 
  6, Joliot Curie, 141980 Dubna, Russia \\and \\
$^3$Stepanov Institute of Physics,
68, Nezavisimosti Ave., 220072, Minsk, Belarus}
\vspace{0.5cm}
\abstract{Following the previously developed approach to the calculation of quantum corrections to the effective potential in arbitrary scalar field theories in the leading logarithmic approximation, we extended it to the next-to-leading order. Based on Bogoliubov-Parasiuk-Hepp-Zimmerman renormalization procedure and the Bogoliu-bov-Parasiuk theorem, we  construct  recurrence relations and  renormalization group equations that allow one to sum up the leading and subleading logarithms  in  all orders of perturbation theory. The formalism is applicable to an arbitrary scalar potential, renormalizable or not. To verify the results, we compare them with a renormalizable model treated within the standard renormalization group approach.}
\\
\text{\footnotesize{E-mails: $^a$yaxibbaev@jinr.ru, $^b$kazakovd@theor.jinr.ru, $^c$mukhaeva@theor.jinr.ru,}} \\
\text{\footnotesize{$^d$dtolkachev@jinr.ru}}

\end{center}
\section{Introduction}

The effective potential in quantum field theory is the first term of the derivative expansion of the effective action, which contains no derivatives. It is a generalisation of the classical potential with account for quantum corrections. The effective potential can be applied in various contexts, in particular, it can be used to determine  the ground state of a given field theory \cite{CW,EA,Iliopoulos:1974ur}. Usually, the calculation of the effective potential includes the computation of one-loop corrections and its further improvement with the help of the renormalization group (RG). The latter can effectively be done in renormalizable models but faces with the lack of formalism in the non-renormalizable case. This was precisely the problem which we solved in a series of papers~\cite{we2015,we2019,Kazakov:2022pkc}, where we constructed the renormalization group equations for an arbitrary scalar potential in the leading logarithmic approximation. We then applied the advocated formalism to various types of potentials, which appear in some cosmological models \cite{Starobinsky:1998fr,Inagaki:2014wva}.

In the present paper, we make the next step and consider the RG equation in the next-to-leading approximation. Here we  face some conceptual problems. First of all, in non-renormalizable models the counter terms that are needed to eliminate the UV divergences do not repeat the original Lagrangian and cannot be absorbed into the redefinition of the original couplings. At each step, new operators appear and one faces infinite arbitrariness. Due to these difficulties, non-renormalizable models are claimed to be irrelevant and are commonly treated only in the framework of the effective field theory approach, where consideration is limited to first orders of perturbation theory, which are supposed to work up to the UV-cutoff limit. In this way, non-renormalizable models are treated in cosmology, in the analysis of effective quark interactions, etc. as outlined in Ref. \cite{Donoghue:1994dn}.  We follow an alternative approach and treat the non-renormalizable model as a UV complete theory. In doing so, we do not solve the ambitious  problem of eliminating  infinite arbitrariness but notice that assuming that UV divergences are subtracted in one way or another, the leading logarithmic approximation (LLA) is independent of the subtraction scheme. This observation allows us  to write down the LL RG-equation which is universal \cite{Kazakov:2022pkc}. Solving this equation we sum up the leading logarithms in all orders of perturbation theory just like in renormalizable models. 

The second observation is that even in the renormalizable case, the NLL approximation depends on the subtraction scheme. This scheme dependence is compensated by the choice of coupling, but the compensation holds only up to neglected orders of perturbation theory.
In the non-renormalizable case, there is no single coupling and the compensation  has a more complicated form \cite{Kazakov:2019wce}. Below we try to trace the scheme dependence in the obtained results, but we must admit that we do not have a full solution of this problem.

Having in mind all these observations, we derive the RG equation for the effective potential in the NLL approximation and its solution  for the simplest quartic scalar model. 

The paper is organized as follows. The first section is dedicated to the basic  formalism and the  derivation of recurrence relations based on the Bogoliubov-Parasiuk theorem. In the second section, we briefly discuss some of our previous results concerning the effective potential in the LL approximation. We present a simple rule that allows one to construct the RG equation for the effective potential. In section four,  we calculate the diagrams and derive recurrence relations in the NLL approximation.  In the next section, we study the scheme dependence problem. Then in the last section, we check our results comparing them with the renormalizable case known in the literature. In conclusion, we discuss the general properties of the all-loop effective potential in the NLLA.

\section{Incomplete $\mathcal{R}$-operation and locality condition}\label{sec:R-operation}

Recall some technical details of the $\cal R$-operation. We follow the analysis carried out in Ref. \cite{we2019}, and  remind only the main results needed for calculating  the effective potential. 

The action of the $\cal R$-operation on the divergent graph $G$ is defined as follows:
\begin{equation}
    \mathcal{R}\circ G= (1-\mathcal{K})\mathcal{R}^\prime\circ G,
\end{equation}
where $\cal K$ is the operator that singles out the divergent part and $\cal R^\prime$  is the incomplete $\mathcal{R}$-operation that subtracts divergences in subgraphs. Thus, the $\mathcal{KR}^\prime\circ G$ is the counterterm corresponding to the graph $G$ \cite{BogoliubovBook,Collins}. In general, when using dimensional regularization the action of the  $\cal R^\prime$ operation on an $n$-loop graph  $G_n$ can be written as
\begin{align}
    \mathcal{R}^\prime G_n& =\frac{\mathcal{A}^{(n)}_{n}}{\epsilon^n}(\mu^2)^{n \epsilon}+\frac{\mathcal{A}^{(n)}_{n-1}}{\epsilon^n}(\mu^2)^{(n-1) \epsilon}+\ldots\frac{\mathcal{A}_1^{(n)}}{\epsilon^n}\mu^{2\epsilon} +\nonumber\\&
    +\frac{\mathcal{B}^{(n)}_{n}}{\epsilon^{n-1}}(\mu^2)^{n \epsilon}+\frac{\mathcal{B}^{(n)}_{n-1}}{\epsilon^{n-1}}(\mu^2)^{(n-1) \epsilon}+\ldots\frac{\mathcal{B}_1^{(n)}}{\epsilon^{n-1}}\mu^{2\epsilon}+\\& +\frac{\mathcal{C}^{(n)}_{n}}{\epsilon^{n-2}}(\mu^2)^{n \epsilon}+\frac{\mathcal{C}^{(n)}_{n-1}}{\epsilon^{n-2}}(\mu^2)^{(n-1) \epsilon}+\ldots\frac{\mathcal{C}_1^{(n)}}{\epsilon^{n-2}}\mu^{2\epsilon}+\nonumber\\& +\text{~lower pole terms like   $~\mathcal{D}^{(n)}_k$, $\mathcal{E}^{(n)}_k$, etc.},\nonumber
\end{align}
where the terms  like $\frac{\mathcal{A}^{(n)}_{k}}{\epsilon^n}(\mu^2)^{k\epsilon}$ appear after subtraction of the $(n-k)$-loop counterterm.
The resulting expression has to be local and should not contain terms like $\log^k(\mu^2)/\epsilon^l$ since in the  coordinate space these terms correspond to non-local operators. Elimination of these terms is necessary to preserve the locality of the counterterms. This is the core idea of the Bogoliubov-Parasiuk theorem and BPHZ-procedure (see Refs. \cite{BP,Hepp,Zimmermann} and in the context of dimensional regularization Ref.\cite{Breitenlohner:1975hg, Breitenlohner:1976te}). The locality requirement imposes constraints and leads to the following relations between the singular parts of the counterterms for the leading, subleading and subsubleading pole terms, respectively:
\begin{align}
        \mathcal{A}^{(n)}_n &=(-1)^{n+1}\frac{\mathcal{A}^{(n)}_1}{n};\\
        \mathcal{B}^{(n)}_n &=(-1)^n ~2! \left(\frac{\mathcal{B}^{(n)}_1}{n}+\frac{\mathcal{B}^{(n)}_2}{n(n-1)}\right);\\
        \mathcal{C}^{(n)}_n&=(-1)^{n+1}~ 3! \left(\frac{\mathcal{C}^{(n)}_1}{2n}+\frac{\mathcal{C}^{(n)}_2}{n(n-1)}+\frac{\mathcal{C}^{(n)}_3}{n(n-1)(n-2)}\right), \mathrm{etc.} \label{RRpoles}
    \end{align}
Using these relations, one can reduce the calculation of the poles in any order of PT to the calculation of  one-, two- and three-loop diagrams, for leading, subleading and subsubleading pole terms, etc.

It is useful also to write down the local expression for the ${\cal KR'}$ terms (counterterms) which actually enter into the recurrence relations. They are equal to
\beq
{\cal KR'}G_n=\sum_{k=1}^n \left(\frac{\mathcal{A}_k^{(n)}}{\epsilon^n} +\frac{\mathcal{B}_k^{(n)}}{\epsilon^{n-1}}+\frac{\mathcal{C}_k^{(n)}}{\epsilon^{n-2}}\right)\equiv
\frac{\mathcal{A}_n^{(n)'}}{\epsilon^n}+\frac{\mathcal{B}_n^{(n)'}}{\epsilon^{n-1}}+\frac{\mathcal{C}_n^{(n)'}}{\epsilon^{n-2}}+\ldots.
\eeq
Then one has, respectively,
\begin{align}
  \mathcal{A}_n^{(n)'}&=(-1)^{n+1}\mathcal{A}_n^{(n)}=\frac{\mathcal{A}_1^{(n)}}{n}, \label{rel2-1} \\
\mathcal{B}_n^{(n)'}&= \left(\frac{2}{n(n-1)} \mathcal{B}_2^{(n)}+\frac{2}{n}\mathcal{B}_1^{(n)}\right) \label{rel2-2},\\
\mathcal{C}_n^{(n)'}&=\left(\frac{2}{(n-1)(n-2)}\frac{3}{n}\mathcal{C}_3^{(n)}+\frac{2}{n-1}\frac{3}{n}\mathcal{C}_2^{(n)}+\frac{3}{n}\mathcal{C}_1^{(n)}\right), ~~\text{etc.}
\label{rel2-3}
\end{align}

One can also write down the corresponding expressions for the coefficients of the logarithms which do not coincide with the coefficients of the poles in subleadfng orders. We denote them by $\bar{\mathcal{A}}_n^{(n)}$, $\bar{\mathcal{B}}_n^{(n)}$ and $\bar{\mathcal{C}}_n^{(n)}$, etc. Then, one has
\beqa
  \bar{\mathcal{A}}_n^{(n)}&=& \mathcal{A}^{(n)}_n= (-1)^{n+1}\frac{\mathcal{A}^{(n)}_1}{n}, \nonumber\\
  \bar{\mathcal{B}}^{(n)}_n&=& (-1)^{n} \left(\mathcal{B}_1^{(n)}+\frac{2}{(n-1)}\mathcal{B}_2^{(n)}\right) \label{Blog}\\
\bar{\mathcal{C}}^{(n)}_n&=& (-1)^{n+1} \left(\frac{n-1}{2}\mathcal{C}_1^{(n)}+2\mathcal{C}_2^{(n)}+\frac{3}{n-2}\mathcal{C}^{(n)}_3\right),~~ \text{etc.} 
 \nonumber
\eeqa

In what follows, we use these relations to extract the logarithmic behaviour of the effective potential in the leading and subleading order.

\section{Effective potential in the LLA}
\label{sec:eff_pot}

\begin{figure}[ht]
 \begin{center}
  \epsfxsize=12cm
 \epsffile{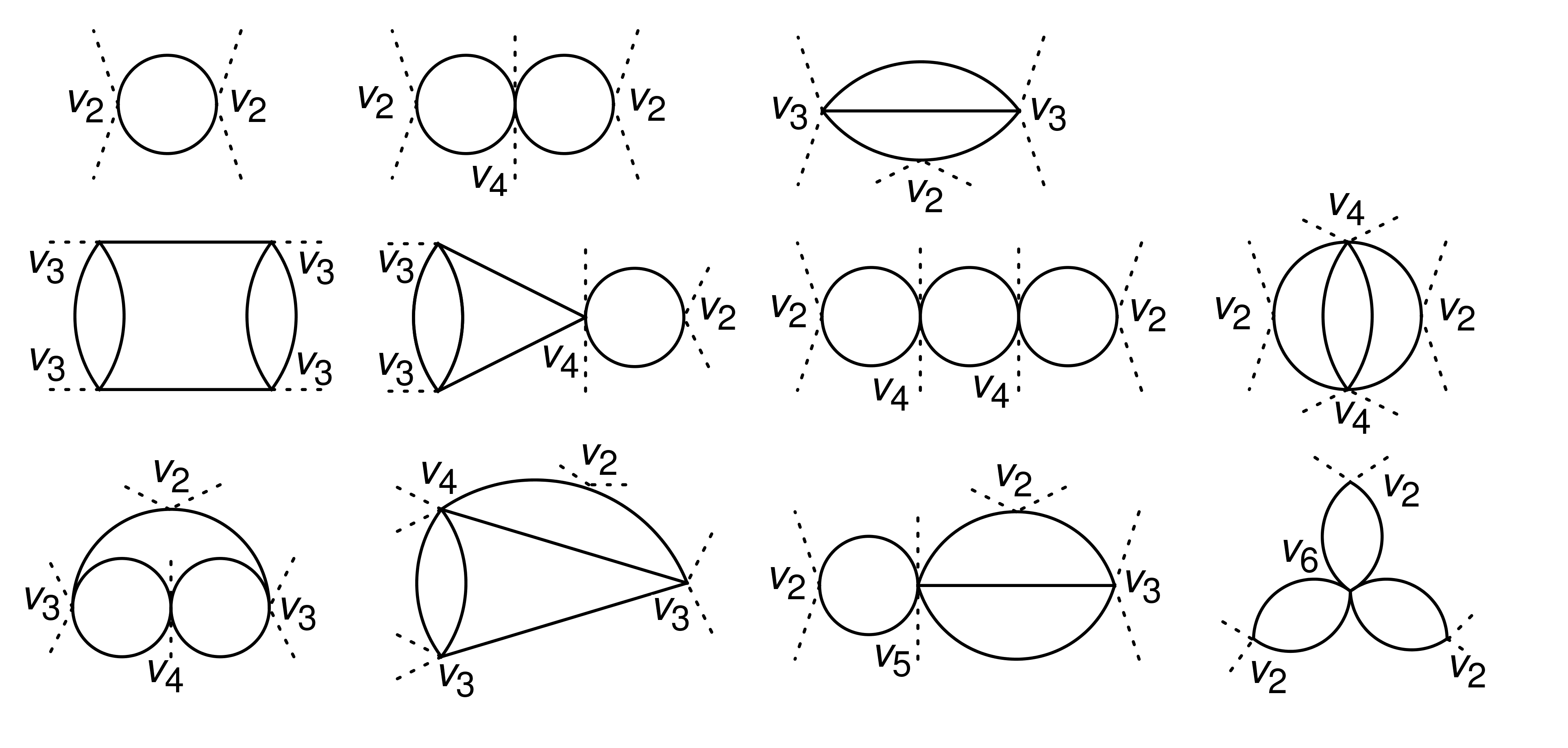}
 \end{center}
 \vspace{-0.4cm}
 \caption{The one-, two- and three-loop vacuum graphs contributing to the effective potential} 
\label{vacgr}
 \end{figure}

Consider the scalar model with an arbitrary potential with a single coupling\cite{EA}
\begin{equation}
    \mathcal{L}=\frac{1}{2}(\partial_\mu \phi)^2-g V_0(\phi).
\end{equation}
As already mentioned, the effective potential is part of the effective action that contains no derivatives of the fields. 
A direct way to calculate the effective potential  perturbatively  is to sum over all 1PI vacuum diagrams acquired using the Feynman rules derived from the shifted action $S[\phi+\widehat \phi]$, where $\phi$ is the classical field obeying the equation of motion and $\widehat \phi(x)$ is the quantum field~\cite{EA,CW}. This results in the propagator of the quantum field  containing  an infinite number of insertions 
of $v_2(\phi)\equiv \frac{d^2V_0(\phi)}{d\phi^2}$, which
acts like a mass term that depends on the field $\phi$: $m^2(\phi)=gv_2(\phi)$, and the vertices generated by expansion of $V_0(\phi+\widehat \phi)$ over $\widehat\phi$.  Then the  effective potential is given in the form of power series over the coupling $g$
\beq
V_{eff}=g\sum_{n=0}^\infty (-g)^n V_n. 
\eeq
The one-, two- and three-loop vacuum diagrams contributing to the effective potential are presented in Fig. \ref{vacgr}, where the classical external fields are symbolically shown by the dotted lines. The number of external lines in each vertex is not fixed and depends on the choice of the classical potential, namely, it is given by the derivatives of the potential $v_n\equiv d^n V_0/d\phi^n$. The number of derivatives  in  its turn corresponds to the number of internal quantum lines in each vertex, as is shown explicitly in Fig. \ref{vacgr}.  
 
 Note, however, that one cannot consider the effective potential  independently of the effective action as a whole. The other terms of the effective action with the derivatives can also contribute to divergent part of  effective potential through the possible divergent subgraphs.
Fortunately, this does not take place in the leading order where one has a self-consistent  equation for the effective potential. However, already in the subleading order one has to add the scalar field propagator graph that contains two derivatives.  We take it into account when considering the subleading expression for the effective potential below.

\begin{figure}[ht]
 \begin{center}
  \epsfxsize=11cm
 \epsffile{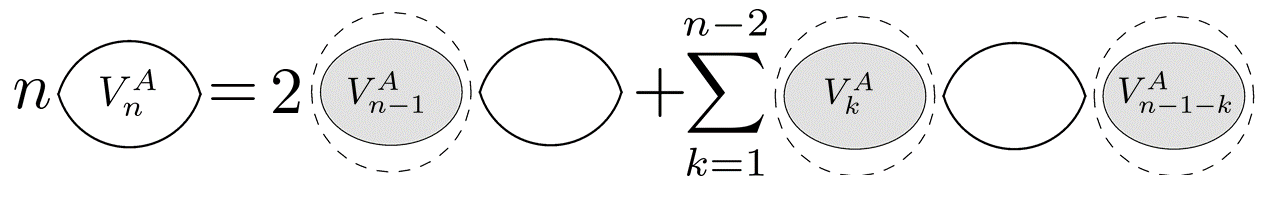}
 \end{center}
 \vspace{-0.4cm}
 \caption{Recurrence relation of the effective potential in the leading order. Grey blobs denote the ${\cal KR}'$ part of the  subgraphs shrunken to a point, namely  $V_k^{'A}$.} 
\label{Rop1loop}
 \end{figure}

The next step is to  evaluate the leading $1/\epsilon^n$ pole terms, $A_n^{(n)}$,  in each diagram in the $n$-th order of PT.  This can be done with the help of eq.~\eqref{rel2-1}-\eqref{rel2-3}, according to which one has to take the diagram with one live loop and $(n-1)$-loop counterterm.
Schematically, eq.~\eqref{rel2-1}-\eqref{rel2-3} is shown in Fig.~\ref{Rop1loop}, where we denoted by $V_k^A$ the leading pole term of  the effective potential.
Strictly speaking, this should be the counterterm  $V_k^{'A}$ according to eq.~\eqref{rel2-1}-\eqref{rel2-3},
however, below we omit the prime symbol to simplify the notation.

\begin{figure}[ht]
 \begin{center}
  \epsfxsize=8cm
 \epsffile{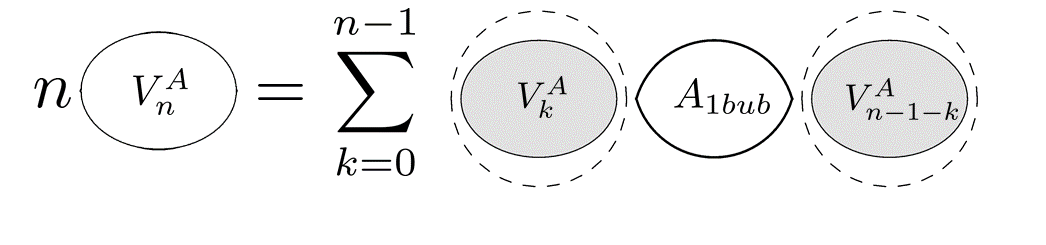}
 \end{center}
 \vspace{-0.4cm}
 \caption{The linear and nonlinear parts of the recurrence relation can be unified in the structure that reminds the one-loop diagram } 
\label{Rop1rule}
 \end{figure}

 The first term in this figure refers to the case when the one-loop diagram is situated at the edge of the diagram and the second term refers to the case when it is in the middle. In fact the first term can be absorbed into the second one modifying the summation limits, as shown in Fig.~\ref{Rop1rule}, which will be useful later.

Following Fig. \ref{Rop1rule} and   eq.~\eqref{rel2-1}, one can write down  a recurrence relation connecting the leading divergences of the $n$-th and the $(n-1)$-th  orders as
\begin{equation}
   n V^A_n=-\frac 14\sum_{k=1}^{n-1} D_2V^A_k D_2 V^A_{n-k-1}, \label{recurrLLA}
\end{equation}
where the derivative $D_2=\frac{\partial^2}{\partial\phi^2}$  comes from  two internal quantum lines and  the factor $-\frac 14$ is the  divergence  of the one loop diagram.  

Introducing  the function 
$$\Sigma_A(z,\phi)=\sum_{n=0}^\infty(-z)^n V^A_n,$$ where $z=g/\epsilon$, one can convert the recurrence relation (\ref{recurrLLA}) into the differential equation
\begin{equation}
\frac{\partial}{\partial z} \Sigma_A=-\frac{1}{4} \left(D_2 \Sigma_A\right), ~ \Sigma_A(0,\phi)=V_0. \label{SigmaAEq}
\end{equation}

Since the coefficients of the leading poles and the leading logarithms are the same, one can just substitute  $\log(\mu^2/m^2)$ instead of $1/\epsilon$  into the solution of eq.~\eqref{SigmaAEq} to obtain the effective potential
\begin{equation}
    V_{eff}=g\Sigma_A|_{z\to g\log\frac{\mu^2}{m^2}}\label{RGeffpotLLA}.
\end{equation}

Notice that the diagram in the rhs of Fig.~\ref{Rop1rule} looks like the one-loop one with the external vertex replaced by the counterterms.  The same is true for equation \eqref{SigmaAEq}, it  has a structure identical to a single-loop diagram but instead of the initial classical potential $V_0(\phi)$ after summation of all orders of PT one has the sum of the leading poles $\Sigma_A(z,\phi)$.  
One actually has a substitution
\beq
D_2 V_0 \rightarrow D_2 V_k^A   \  \ \mbox{or}  \  \  D_2 V_0 \rightarrow D_2 \Sigma_A,  \label{subst}
\eeq
valid for the counterterms of a given order $V_k^A$,  or for the total sum  $\Sigma_A $, respectively.
The number of derivatives here depends only on the number of internal propagators outgoing from the vertex.  In the subleading case, we have $D_3$ and $D_4$ as well as $D_2$. We refer to this prescription as the {$\mathcal{R}$}-rule.

This $\mathcal{R}$-rule seems to have a general nature and is directly related to the structure of the $\mathcal{R}'$-operation. To derive  the recurrence relation or the differential equation,  one has to take the corresponding lowest order diagrams and make a replacement like (\ref{subst}) with the number of derivatives equal to the number of internal quantum lines. We check and apply this rule when constructing the corresponding recurrence relation and the differential equation for the subleading divergences in the next Section.

\section{Effective potential in the NLLA}
\label{sec:subl}

To calculate the effective potential for the subleading order, we have to  take into account two-loop diagrams according to eq.~\eqref{rel2-2}. We start with establishing the recurrence relations. Using the  $\mathcal{R}$-rule formulated above, we have the  following set of diagrams,  as shown in Fig.~\ref{B1g2loop}.  Here we adopt the notation $V_k^J$,  where $J=A$ or $J=B$  for the leading and subleading divergences, respectively, and $k$ denotes the number of loops.

\begin{figure}[ht]
 \begin{center}
  \epsfxsize=17.0cm
 \epsffile{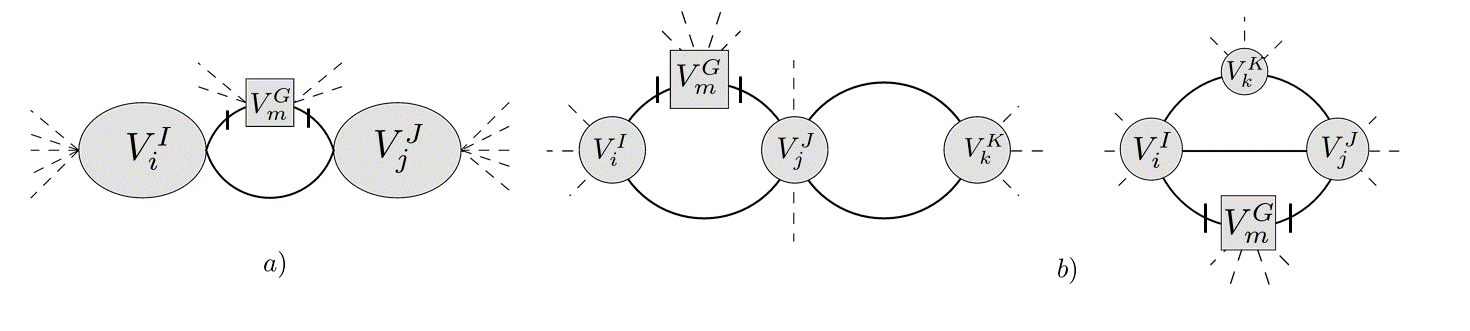}
 \end{center}
 \vspace{-0.2cm}
 \caption{The one- and two-loop diagrams contributing to the recurrence relation for the next-to-leading divergences according to the $\mathcal{R}$-rule. $V_k^G$ corresponds to the part of the effective action with two derivatives (they are denoted as notches on the lines).} 
\label{B1g2loop}
 \end{figure}
 
In order to collect the subleading divergence ($V^B_i$), one should take the subleading terms either from the survived diagram and take the leading contributions from the counterterms, or vise versa, take the leading terms from the survived diagram and subleading contribution from one of the counterterms. In essence, we have an arrangement of $V^A_i$ and $V^B_j$  in each term, where the index $B$ occurs only once. Notice that the contribution from the effective action $V_k^G$ is also subleading.

Turning back to eq.~\eqref{rel2-1}-\eqref{rel2-3}, we have the following contribution from the one-loop diagram, shown in Fig.~\ref{B1g2loop}, where we explicitly show the leading and subleading contributions from each subgraph.
\begin{figure}[ht]
 \begin{center}
  \epsfxsize=12.0cm
 \epsffile{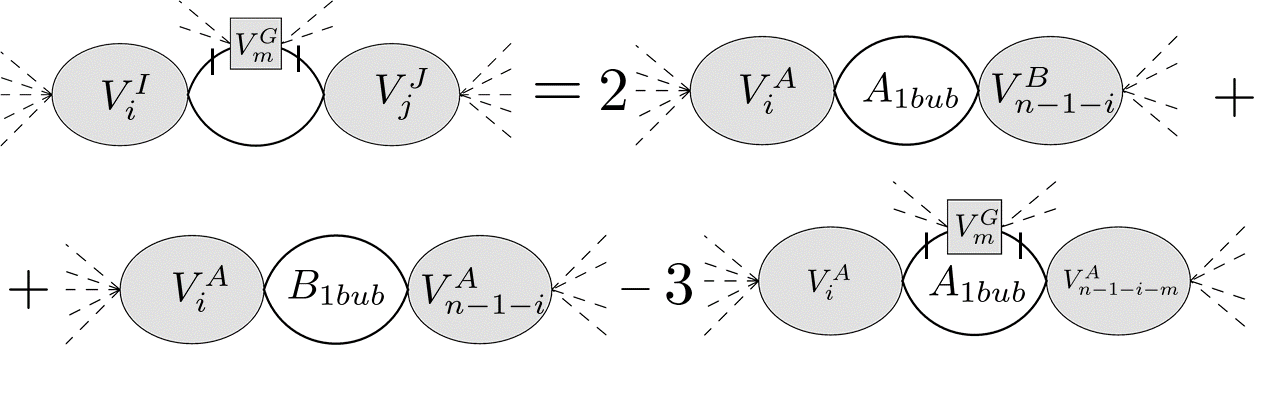}
 \end{center}
 \vspace{-0.2cm}
 \caption{Part of the recurrence relation contributing to the subleading order containing the one-loop survived diagram.} 
\label{B1g2loop}
 \end{figure}
 
 To write down an explicit expression, we need the leading and subleading contribution from the one-loop diagram
 \begin{equation}
   V_{1bub}= \frac{1}{4\epsilon} +\frac{c_1}{4}.
\end{equation}
Here $c_1$ is the subtraction scheme dependent constant.

As a result, an analytical contribution from one-loop diagrams looks like
\begin{equation}
    \begin{split}
        &\sum_{k=0}^{n-1}2 \frac 14 D_2 V^B_k D_2 V^A_{n-k-1}- 3 \frac 14\sum_{l,k=0}^{l+k<n-2} D_2 V^A_l D_2 V^A_k D_2 V^G_{2, n-l-k-1}.
    \end{split}
\end{equation}

Now we consider the contribution from the two-loop diagrams. We have two types of contributions: the two-loop bubble-type and the sunset-type, as shown in Fig.~\ref{B1g2loop}. For this purpose, we need the explicit expression for their divergent parts. One has
\beqa
 V_{2bub}&=&v_2^2 v_4 \left(\frac{1}{8 \epsilon^2} +\frac{c_1}{4 \epsilon}\right),\\
   V_{2sun}&=&v_2^2 v_4 \left(\frac{1}{8 \epsilon^2} -\frac{1+c_1}{8 \epsilon}\right).
\eeqa
Here again $c_1$ is the one-loop subtraction constant, but we prefer to procceed in the minimal subtraction scheme $c_1=0$, for the sake of simplicity. We discuss the scheme dependence in the subleading order in the next Section.

As a result, one has the following analytical contribution to the recurrence relation coming from by the double bubble contribution according to the Fig. \ref{B1g2loop}b):
\begin{equation}
    \begin{split}
        &- \sum_{l,k=0}^{l+k<n-1} 
       \left[{ \frac 18 D_2 V^A_l D_2 V^A_k D_4 V^B_{n-l-k-2}
        + 2\frac 18 D_2 V^A_l D_2 V^B_k D_4 V^A_{n-l-k-2}}\right]\\
        &+ 3 \frac 18\sum_{l,k,m=0}^{l+k+m<n-3} {  D_2 V^A_l D_2 V^A_k D_4 V^A_m D_2 V^G_{2, n-l-k-m-2}}.
    \end{split} 
    \label{dbb}
\end{equation}
In this expression again we  keep only one term of the subleading order, thus the index $B$ or $G$ can appear only once. Note, that we take the leading pole from the survived diagram in eq.~\eqref{dbb}.

The contribution of the sunset diagram is as follows:
\begin{equation}
    \begin{split}
        &- \sum_{l,k=0}^{l+k<n-1} \left[
        \frac 18 D_3 V^A_l D_3 V^A_k D_2 V^B_{n-l-k-2} 
        +2 \frac 18 D_3 V^A_l D_3 V^B_k D_2 V^A_{n-l-k-2} \right. \\
        & \left. - \frac{1}{8} D_3 V^A_l D_3 V^A_k D_2 V^A_{n-l-k-2}+ 2\frac 18\sum_{l,k,m=0}^{l+k+m<n-3}  D_3 V^A_l D_3 V^A_k D_2 V^A_m D_2 V^G_{2, n-l-k-m-2} \right].
    \end{split}
\end{equation}
Again, the logic here is the same: one  should keep only one subleading contribution from each shrunken or survived subgraph.

Summing up all the contributions, one can write down the recurrence relation~\eqref{rel2-1}-\eqref{rel2-3} in the NLLA as
\beqa
&&n(n-1){V}_n^{B}=\nonumber  \\
&& (n-1)\left(2 \frac 14\sum_{k=0}^{n-1}  D_2 V^B_k D_2 V^A_{n-k-1} - 3 \frac 14\sum_{l,k=0}^{l+k<n-2} D_2 V^A_l D_2 V^A_k D_2 V^G_{2, n-l-k-1}\right)\nonumber\\
 && - \sum_{l,k=0}^{l+k<n-1} 
       \left[{ \frac 18 D_2 V^A_l D_2 V^A_k D_4 V^B_{n-l-k-2}
        + 2\frac 18 D_2 V^A_l D_2 V^B_k D_4 V^A_{n-l-k-2}}\right]\nonumber\\
       & &+  3 \frac 18\sum_{l,k,m=0}^{l+k+m<n-3} { D_2 V^A_l D_2 V^A_k D_4 V^A_m D_2 V^G_{2, n-l-k-m-2}} \nonumber \\
         &&- \sum_{l,k=0}^{l+k<n-1} \left[
        \frac 18 D_3 V^A_l D_3 V^A_k D_2 V^B_{n-l-k-2} 
        +2 \frac 18 D_3 V^A_l D_3 V^B_k D_2 V^A_{n-l-k-2\nonumber} \right.\\
        &&- \left. \frac{1}{8} D_3 V^A_l D_3 V^A_k D_2 V^A_{n-l-k-2} \right]+\nonumber\\
        &&+ 2\frac 18\sum_{l,k,m=0}^{l+k+m<n-3} 
         D_3 V^A_l D_3 V^A_k D_2 V^A_m D_2 V^G_{2, n-l-k-m-2} \label{recrel1}.
\eeqa

\begin{figure}[ht]
 \begin{center}
  \epsfxsize=5.2cm
 \epsffile{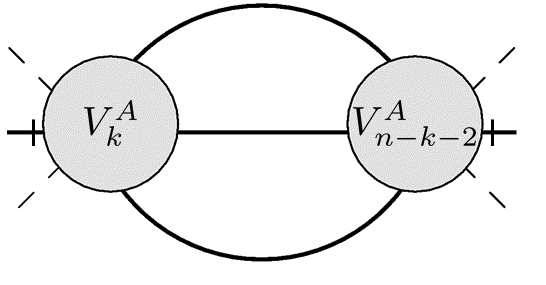}
 \end{center}
 \vspace{-0.2cm}
 \caption{Propagator type diagram giving a contribution to the part of the effective action with two derivatives.} 
\label{g2Roploop}
 \end{figure}

\begin{figure}[ht]
 \begin{center}
  \epsfxsize=3.2cm
 \epsffile{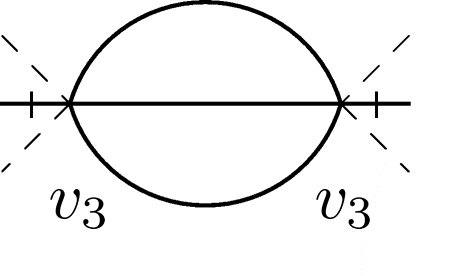}
 \end{center}
 \vspace{-0.2cm}
 \caption{Two-loop walnut-type diagram.} 
\label{g2loop}
 \end{figure}

As  already mentioned, contrary to the LLA in the next-to-leading  order we have to take into account the part of the effective action with two derivatives which we denoted by $V^G$.  Therefore, we need to write down the corresponding recurrence relation for the 
propagator type diagram shown in Fig.~\ref{g2Roploop}. 
 This recurrence relation follows from the $\cal R$-rule  and reads
\begin{equation}
    n(n-1)V^G_n= -\frac{1}{24}\sum_{k=0}^{n-2} D_3 V^A_k D_3 V^A_{n-k-2}\label{recrel2},
\end{equation}
where the coefficient $-\frac{1}{24}$ is nothing else than the single pole contribution of the two-loop walnut-type diagram shown in Fig.~\ref{g2loop}.
Despite the visual complexity of these recurrence relations arising from the $\cal R$-operation, there is a fairly simple structure behind them, which, as mentioned above, repeats the structure of loop contributions.

By multiplying the recurrence relations~\eqref{recrel1}-\eqref{recrel2}  by $(-z)^{n}$ and summing up, we can obtain the RG equations for the functions $\Sigma_B=g\sum_{n=0}^\infty (-z)^n V^B_n$ and  $\Sigma_G=g\sum_{n=0}^\infty  (-z)^nV^G_n$. Namely, in the minimal subtraction scheme one gets:
 \beqa
     &&   \frac{\partial^2}{\partial z^2}\Sigma_B=  \nonumber \\
     &&-2\frac{1}{4}\frac{\partial}{\partial z}\left\{ 2 D_2 \Sigma_A D_2 \Sigma_B-\left(D_2 \Sigma_A\right)^2 D_2 \Sigma_G\right\}-\nonumber \\&&-2 \left\{\frac{1}{8} (D_3 \Sigma_A)^2 D_2 \Sigma_B-2  \frac{1}{8} (D_3\Sigma_A)^2 D_2 \Sigma_A D_2 \Sigma_G+\right.\nonumber \\ &&\left. +2 \frac{1}{8} D_3 \Sigma_A D_3 \Sigma_B D_2 \Sigma_A-\frac{1}{8} (D_3 \Sigma_A)^2 D_2 \Sigma_A+\frac{1}{8}(D_2 \Sigma_A)^2 D_4 \Sigma_B -\right.\nonumber  \\ &&\left. -2 \frac{1}{8}(D_2\Sigma_A)^2 D_4 \Sigma_A D_2 \Sigma_G + 2\frac{1}{8}D_2\Sigma_A  D_2\Sigma_B D_4\Sigma_A  \right\},
   \label{SigmaBeq} \\
   &&  \frac{\partial^2}{\partial z^2}\Sigma_G=\frac{2}{24}(D_3 \Sigma_A)^2. \label{SigmaGeq}
\eeqa

Here we intentionally left all combinatorial coefficients and coefficients of the survived diagrams to demonstrate similarities between the recurrence relations and the obtained equations. 

Despite the fact that these equations appear to be extremely complicated, their form reproduces the structure of the diagrams, following the arrangement of contributions according to the degree of divergence: a single-loop diagram contributes only to the leading pole and can produce scheme dependent parts in the subleading order, two-loop diagrams contribute to both the leading and the subleading poles, and the propagator correction always contributes to the subleading pole. To obtain the contribution of the subleading pole, it is necessary to multiply the set of leading poles by at least one subleading pole. 

To obtain the effective potential in the subleading order, one first needs to solve the equations  for $\Sigma_A, \Sigma_B$ and $\Sigma_G$ and to use the relation~\eqref{Blog}. This relation can also be converted into the differential equation for the 
function $\bar{\Sigma}_B$ which is a function that sums all logarithmic corrections, i.e., the effective potential in the subleading order.  The corresponding equation takes the form
\begin{equation}
\begin{gathered}
    \frac{\partial}{\partial z}\bar{\Sigma}_B= \frac{\partial}{\partial z}\left\{ 2 D_2 \Sigma_A D_2 \Sigma_B-\left(D_2 \Sigma_A\right)^2 D_2 \Sigma_G\right\}-\\-2 \left\{\frac{1}{8} (D_3 \Sigma_A)^2 D_2 \Sigma_B-2  \frac{1}{8} (D_3\Sigma_A)^2 D_2 \Sigma_A D_2 \Sigma_G+\right.\\ \left. +2 \frac{1}{8} D_3 \Sigma_A D_3 \Sigma_B D_2 \Sigma_A-\frac{1}{8} (D_3 \Sigma_A)^2 D_2 \Sigma_A+\frac{1}{8}(D_2 \Sigma_A)^2 D_4 \Sigma_B -\right. \\ \left. -2 \frac{1}{8}(D_2\Sigma_A)^2 D_4 \Sigma_A D_2 \Sigma_G + 2\frac{1}{8}D_2\Sigma_A  D_2\Sigma_B D_4\Sigma_A  \right\}.
\end{gathered}\label{SigmaBlog}
\end{equation}
This is the first order differential equation with the known rhs. After integration one should replace $z \rightarrow g \log(\mu^2/m^2(\phi))$ to obtain the NLLA part of the effective potential where the field dependent mass $m^2(\phi)=g v_2(\phi)$. Notice that while the order of the differential equations for the corrections in subsequent orders of PT  increases at each step, the  differential equations for the logarithms are always linear. 

\section{Subtraction scheme dependence}\label{sec:scheme}

In this section we discuss the subtraction scheme dependence of the results.  While  the leading divergences (leading poles) are scheme independent, already the subleading poles depend on the subtraction procedure and so do all the subsequent ones. This scheme dependence is also present in renormalizable theories and is absorbed into the  finite redefinition of the coupling. And though the compensation of the scheme dependence in perturbation theory is not exact, it is valid up to discarded terms of PT. 

In non-renormalizable theories the situation is slightly different.  As  already mentioned, in nonrenormalizable theories the structure of the counterterms does not repeat the initial Lagrangian. So the scheme dependence is not compensated by the redefinition of a single coupling but requires the redefinition of new emergent operators. Still, in general, in a full theory, the scheme  dependence has to be eliminated. Here we discuss some elements of this procedure.

In our earlier works \cite{Kazakov:2017xul,Kazakov:2019heg}, we considered the maximally supersymmetric Yang-Mills theories in $D=6, 8, 10$ dimensions and noticed that the scheme dependence 
of subleading divergences of a particular set of diagrams, the so-called ladder  diagrams, is described by the redefinition of the coupling $z=g/\epsilon$ given by the following substitution:
\begin{equation}
    z \rightarrow z(1 + \epsilon ~c_1), \label{z}
\end{equation}
where $c_1$ is the arbitrary coefficient that appears in subtracting  the one-loop diagram and characterises the choice of the scheme
\beq
1/\epsilon+c_1.
\eeq
The minimal subtraction scheme corresponds to $c_1=0$.

In the case of the effective potential,  the situation is similar. The transformation to a non-minimal scheme results in the appearance of an additional term in the solution of eq.~\eqref{SigmaBeq}, i.e., one has
\beq
\Sigma_B^{c_1}(z,\phi)=\Sigma_B^{\mathrm{MS}}+c_1\Delta\Sigma_B, \label{extra}
\eeq
where the first term is the sum of the subleading counterterms in the minimal scheme  and the second term is the addition that appears when the subtraction scheme is changed.

Taking into account the structure of the $\cal R$-operation, namely, the linearity of eq.~\eqref{SigmaBeq} with respect to $\Sigma_B$, an additional term in a non-minimal scheme can only enter linearly in $c_1$. Making a substitution (\ref{extra}) in eq.~\eqref{SigmaBeq}, we obtain  the equation for the subleading counterterms  in an arbitrary non-minimal scheme (hereafter we omit the MS label for simplicity of notation)
\begin{align}
& \frac{\partial^2}{\partial z^2}\Sigma_B+c_1 \frac{\partial^2}{\partial z^2}\Delta\Sigma_B=\\
 &      -\frac{1}{2}\frac{\partial}{\partial z}\left\{ 2 D_2 \Sigma_A D_2 \Sigma_B-\left(D_2 \Sigma_A\right)^2 D_2 \Sigma_G\right\}-\nonumber \\&
       - \left\{\frac{1}{4} (D_3 \Sigma_A)^2 D_2 \Sigma_B-2  \frac{1}{8} (D_3\Sigma_A)^2 D_2 \Sigma_A D_2 \Sigma_G+
       \right.{}\nonumber\\ 
			& \left.
       + \frac{1}{4} D_3 \Sigma_A D_3 \Sigma_B D_2 \Sigma_A-\frac{1}{8} (D_3 \Sigma_A)^2 D_2 \Sigma_A+\frac{1}{8}(D_2 \Sigma_A)^2 D_4 \Sigma_B -
       \right.{}\nonumber\\ 
			& \left.
       - \frac{1}{4}(D_2\Sigma_A)^2 D_4 \Sigma_A D_2 \Sigma_G + \frac{1}{4}D_2\Sigma_A  D_2\Sigma_B D_4\Sigma_A  \right\}
       +
       \nonumber\\ &+\frac{1}{4} c_1
  \left\{(D_2\Sigma_A)^2 \left(-D_4\Delta \Sigma_B\right)
   - 2 D_3\Sigma_A D_2\Sigma_A D_3\Delta \Sigma_B-2 D_4\Sigma_A D_2\Sigma_A D_2\Delta \Sigma_B-
 \right.{}\nonumber\\ 
			& \left.
   - 4 D_2\Sigma_A D_2 \frac{\partial}{\partial z}\Delta \Sigma_B-\left((D_3\Sigma_A)^2+4 D_2 \frac{\partial}{\partial z}\Sigma_A\right) D_2\Delta \Sigma_B\right\}.
        \label{eq:sabl_MSbar_2}
    \end{align}  

In ~\eqref{eq:sabl_MSbar_2} we can separate the equation in the  minimal scheme part and a new additional term proportional to $c_1$ corresponding to a non-minimal scheme. Then the part proportional to $c_1$ should be zero. This gives the equation for $\Delta \Sigma_B$:
\begin{align}
&\frac{\partial^2}{\partial z^2}\Delta\Sigma_B= 
  (D_2\Sigma_A)^2 \left(-D_4\Delta\Sigma_B\right)
   - 2 D_3\Sigma_A D_2\Sigma_A D_3\Delta\Sigma_B-2 D_4\Sigma_A D_2\Sigma_A -
 {}\nonumber\\ 
			& 
   - 4 D_2\Sigma_A D_2\frac{\partial}{\partial z}\Delta\Sigma_B-\left((D_3\Sigma_A)^2+4 D_2 \frac{\partial}{\partial z}\Sigma_A\right) D_2\Delta\Sigma_B.
    \label{scheme}
\end{align}
To simplify  eq.~\eqref{scheme},  one can use eq.~\eqref{SigmaAEq} for $\Sigma_A$. Differentiating it twice with respect to $\phi$, one gets 
$ D_2\frac{\partial}{\partial z} \Sigma_A \to -\frac{1}{2} \left(D_3\Sigma_A\right)^2 -\frac{1}{2}D_2\Sigma_A D_4\Sigma_A$. Using this relation, equation for $\Delta\Sigma_B$ takes the form
\begin{align}
&\frac{\partial^2}{\partial^2 z}\Delta\Sigma_B= \label{dd_eq}\\
    & \frac{1}{4}\left(D_3\Sigma_A\right)^2 D_2\Delta\Sigma_B-\frac{1}{4} D_3\Sigma_A D_3 \Delta\Sigma_B-\frac{1}{4}D_4\Sigma_A D_2\Delta\Sigma_B-D_2\Sigma_A D_2 \frac{\partial}{\partial z}\Delta\Sigma_B.
    \nonumber
\end{align}
A direct solution to this equation is difficult. However, having in mind the analogy with the earlier considered case \cite{Kazakov:2017xul,Kazakov:2019heg}, one can check that the following solution takes place:
\begin{align}
       \Delta\Sigma_B(z,\phi) = \epsilon z \frac{\partial}{\partial z}\Sigma_A(z,\phi),
        \label{eq:MSbar_add}
\end{align}
with arbitrary normalization.
Thus, the effective potential in the leading and subleading order in an arbitrary subtraction scheme has the form
\beq
\Sigma(z,\phi) = \Sigma_A(z,\phi)+\epsilon z \Sigma_B(z,\phi)+\epsilon z c_1 \frac{\partial}{\partial z}\Sigma_A(z,\phi),
\label{sum}
\eeq
where the functions $\Sigma_A$ and $\Sigma_B$ satisfy eqs (\ref{SigmaAEq}) and (\ref{SigmaBeq}), respectively, and we put an extra $\epsilon$ factor in the subleading order.

One can see that eq.(\ref{sum}) is consistent with the substitution (\ref{z}), which we earlier observed in 
the case of scattering amplitudes in gauge theories. Apparently, this is a common feature that reflects the structure of the $\mathcal{R}$-operation. 

So far, the scheme dependence described above reflected arbitrariness in the subtraction of a one-loop graph. Subtraction of higher order graphs is not that straightforward and requires further investigation.

\section{Verification of the results}\label{sec:examples}

Clearly, eqs. \eqref{SigmaBeq} and \eqref{SigmaGeq} are difficult to analyze in the general case. To see some common features and verify our results, we consider some simple examples.

\subsection{General power-like potential}

We start with  the general power-like potential that we  considered in the LLA in \cite{Kazakov:2022pkc}

\begin{equation}
    V_0=\frac{\phi^p}{p!}
\end{equation}
We use the following ansatz for the solutions of the RG-equations:
\begin{equation}
    \Sigma_A=\frac{\phi^p}{p!} f_A(y), ~ \Sigma_B=g\frac{\phi^p}{p!} f_B(y),~\Sigma_G=g\phi^2 f_G(y),
\end{equation}
where the dimensionless variable $y= z \phi^{p-4}$ is introduced.
Then, the RG equation in the LLA takes the following form: 
\begin{equation}
    p! f_A'=-\frac{1}{4 }\left( \left((p-4)^2 y^2 f_A''+(p-4)(3 p-5) y f_A'\right)+(p-1) p f_A\right)^2,
\end{equation}
and the NLLA equation reads
\begin{align}
    p! f_B''&= a_5(y) f^{(4)}_B+a_4(y) f^{(3)}_B+a_3(y)f''_B+a_2(y)  f'_B+a_1(y) f_B+a_0(y),\nonumber\\
       -12 (p!)^2 f''_G&=\left((p\!-\!4)^3 y^3 f_A^{(3)}\!+\!3 (p\!-\!4)^2 (2p\!-\!5) y^2 f_A' \right. \nonumber\\
       &\left.+ \!(p\!-\!4) (p\!-\!2) (7p\!-\!15) y f_A' 
       +\!(p\!-\!2) (p\!-\!1) p f_A\right)^2 ,
\label{SubleadVp}
\end{align}
where the coefficients $a_i$ are equal to:
\begin{align}
a_5& = (p-4)^4 y^4 f_A'(y),\nonumber\\
a_4& =(p-4)^2 y^2 \left((p-4) (9 p-29) y f_A'(y)-(p-1) p f_A(y)\right),\nonumber\\
a_3& =(p-4)^2 y^2 \left(\frac{(p-4)^2 y^2 f_A''(y){}^2}{4f_A'(y)}+ (p-2) (9 p-22) f_A'(y)+\right.{}\nonumber\\ 
			&\left.+ (p-4) y \left(2 (p-2) f_A''(y)-(p-4) y f_A^{(3)}(y)\right)\right),\nonumber\\
a_2& = \left( (p-4) y \left((p-2) (3 (p-10) p+55) f_A'(y)+\right.\right.{}\nonumber\\ 
			& \left.\left. +(p-4) (3 p-5) y \left((5 p-13) f_A''(y)+(p-4) y f_A{}^{(3)}(y)\right)\right)-\right.{}\nonumber\\ 
			&\left. -\frac{(p-4) (3 p-5) y f_A''(y)}{4f_A'(y)} \left((p-4) y \left(2 (3 p-5) f_A'(y)+3 (p-4) y f_A''(y)\right)+2 (p-1) p f_A(y)\right)+ \right.{}\nonumber\\ 
			&\left.+ (p-4) (p-2) (p (23 p-97)+100) y f_A'(y)-8 (p-2) (p-1) p (2 p-5) f_A(y)\right), \nonumber\\  
a_1& = -\frac{(p-1) p f''(y)}{4 f_A'(y)} \left((p-4) y \left(2 (3 p-5) f_A'(y)+3 (p-4) y f_A''(y)\right)+2 (p-1) p f_A(y)\right)+\nonumber\\&+3 (p-2) (p-1) p (2 p-5) f_A'(y)+(p-4) (p-1) p y \left((p-4) y f^{(3)}(y)+(5 p-13) f''(y)\right),\nonumber\\
a_0&=\frac{1}{4 f_A'(y)}\left(2 (p-2) f_A'(y)+(p-4) y f_A''(y)\right){}^2 \left((p-4) y \left((3 p-5) f_A'(y)+(p-4) y f_A''(y)\right)+\right.{}\nonumber\\ 
			&\left. +(p-1) p f_A(y)\right),            
\end{align}
and primes represent derivatives over $y$ as $f'= \dfrac{d}{d y}f(y)$ and the same for the higher derivative terms.

Equations (\ref{SubleadVp}) are the ordinary differential equations. Notice that in the subleading order one has the second order equation. 
This is a general feature: the order of the equation increases with  order of approximation; one has the $(n+1)$-th order ODE in the $\mathrm{N^nLLA}$.

\subsection{Renormalizable model}

To verify our results in analytical form, we consider the renormalizable case  and put $p=4$ in the above equations. Then in the LLA equation one has
\begin{equation}
    f'_A=-\frac{3}{2}f_A^2, ~f_A(0)=1, \label{Aeqp4}
\end{equation}
and its solution is  a geometric progression 
\begin{equation}
    f_A(y)=\frac{1}{1+\frac{3}{2}y}, \label{fa}
\end{equation}
where one can recognize the first coefficient of the beta-function in the quartic scalar model\cite{CW,Kazakov:2022pkc}. This solution is known to be characterised by the Landau pole.

The NLLA equation is also  simplified 
\begin{equation}
   f_B''+6 f_A f_B' +\frac{3}{2}f_B \left(4f_A'+9 f_A^2\right)=3 f_A^3 \left(1+4 \frac{f_A'}{f_A^2} f_G +2 \frac{f_G'}{f_A}+6f_G\right) \label{Beqp4}
\end{equation}
with the initial conditions $f_B(0)=0,f'_B(0)=0$ according to the PT-expansion.
For the term with the derivatives $f_G$, one obtains
\begin{equation}
     f_G''= \frac{1}{12}f_A^2 \label{Geqp4}
\end{equation}
with the initial conditions $f_G(0)=f'_G(0)=0$.

These equations also have simple analytical solutions

\begin{equation}
    f_G(y)=-\frac{f_A-f_A \log \left(f_A\right)-1}{27 f_A}\label{fGpoles}
\end{equation}
and 
\begin{equation}
   f_B(y)= \frac{2}{27} \left((1-f_A) (1-17 f_A)-16 f_A \log \left({f_A}\right)\right),\label{fBpoles}
\end{equation}
where $f_A$ is given by (\ref{fa}). These solutions are scheme dependent and are written here in the MS-scheme and are completely consistent with the results of Ref. \cite{Kazakov:1979ik}.

Finally, one arrives at the equation for the NNLA contribution to the effective potential according to \eqref{Blog} and \eqref{SigmaBlog}.  Changing the argument $y \rightarrow gl =g \log(\mu^2/m^2(\phi))$, one has
\begin{equation}
    \frac{d}{d l}\bar{f}_{B}(l)=\frac{1}{8}\frac{d}{dl} \left(f_A\left(f_Af_G- f_B\right)\right)-2 f_A^2 \left(-\frac{1}{16} f_A +\frac{9}{32}f_B-\frac{3}{8} f_G f_A\right). \label{fBeff}
\end{equation}

Integrating, one gets
\begin{equation}
    \bar{f}_{B}(l)=\frac{3 g l+34 \log \left(1+\frac{3 g l}{2}\right)}{18 (1+\frac{3}{2}g l)^2}. \label{fBlog}
\end{equation}
In this expression, one can recognise the first coefficients of the beta-function of the $\phi^4$ theory \cite{Kazakov:1979ik,Kastening:1991gv,Chung:1999gi}.  The scheme dependence here is manifested in the $3 gl$ term in the numerator, which is written in the MS scheme and is different otherwise. 

These results can be checked using the Ovsyannikov-Callan-Symanzik equation explicitly.
Recall that the usual RG-equation for the effective potential \cite{CW,EA} is
\begin{equation}
\left(\mu^2\frac{\partial}{\partial \mu^2}+\beta(g)\frac{\partial}{\partial g}-\gamma(g)\right)V_{eff}(l)=0
\end{equation}
with the solution
\begin{equation}
    V_{eff}(l)= \frac{\phi^4}{4!} f(\bar{g}(l)) \exp{\left(\int_{g}^{\bar{g}} d x ~\frac{\gamma(x)}{\beta(x)}\right)},\label{OCSsol}
\end{equation}
where $\bar{g}$ is a solution of the equation
\begin{equation}
    \frac{d}{d l}\bar{g}(l)=\beta(\bar{g}(l))\label{GLeq}
\end{equation}
and $f(g)=g(1+c_1g+...)$ is the normalization function that depends on the scheme.
Using the known expansion of the anomalous dimension $\gamma(g)=\gamma_1 g^2+O(g^3)$ and the beta-function $\beta(g)=\beta_0 g^2+\beta_1 g^3+O(g^4)$, one can get the leading and subleading logarithmic corrections to the effective potential in the MS-scheme

\begin{equation}
    V_{eff}(\phi)= g\frac{\phi^4}{4!} \times \left(\frac{1}{(1+\beta_0 gl)}+ g\frac{2\gamma_1 l-\beta_1/\beta_0 \log(1+\beta_0g l)}{(1+\beta_0 gl)^2}\right) \label{RGllog},
\end{equation}
where for the $\phi^4$-theory one has $\beta_0=\frac32, \beta_1=-\frac{17}{6}, \gamma_1=\frac{1}{12}$. One can check that expression (\ref{RGllog}) precisely coincides  with the leading order  (\ref{fa}) and subleading order (\ref{fBeff}) corrections obtained above. Remind that in the effective potential one should put  $l=\log(\mu^2/m^2(\phi))$.

\section{Summary}

To summarise our findings, we claim that
\begin{itemize}
\item The BPHZ $\mathcal{R}$-operation works in non-renormalizable models equally well as in renormalizable ones.
\item The consequence of locality of the $\mathcal{R}$-operation leads to relations between the counter terms of subsequent orders of PT. The relations allow one to get recurrence relations for the counter terms in the leading, subleading, etc. approximations.
\item These recurrence relations can be converted into the differential equations, which in general are the partial differential equations.
They are the first order DE in the leading approximation and increase in order at each next step.
\item One can also write down the differential equations for the logarithms. In the leading order they coincide with those for the counter terms. In subsequent orders they are different, they are always linear differential equations with the rhs defined by the solutions of  equations for the counter terms, as exemplified in \eqref{SigmaBlog}.  These equations are nothing else but the RG equations that sum up the leading, subleading, etc logarithms in all orders of PT. 
\item We have demonstrated how the advocated approach works  in the LLA and in NLLA. In the last case, like in the renormalizable counterpart, one has a scheme dependence.
\item Application to the renormalizable case  shows total identity of the results with those known from the literature. We would like to stress, however, that in general our approach is based on the requirement of locality of the counter terms and does not refer to the multiplicativity of  renormalization whatsoever.
\item The weak point of our analysis is, of course, the lack of a solution to the problem of infinite arbitrariness that is immanent to non-renormalizable theories. We just stress that the LLA does not depend on it and the NLLA is under control. This, however, presents the main challendge.
\end{itemize}

\subsection*{Acknowledgements}
R.M. Iakhibbaev and D.M. Tolkachev are thankful to CERN-TH for hospitality in the process of preparing this publication. Useful discussions with  M.E. Shaposhnikov, A. L. Kataev and S.V. Mikhaylov are acknowledged. The authors thank V.A. Smirnov for attracting our attention to  Refs.\cite{Breitenlohner:1975hg,Breitenlohner:1976te}.
A. I. Mukhaeva's work is supported by the Foundation for the Advancement of Theoretical Physics and Mathematics BASIS, No 24-1-4-36-1.

\bibliographystyle{unsrt}
\bibliography{refs}

\end{document}